\documentclass[12pt,a4paper]{article}

\usepackage{amsmath}
\usepackage{amssymb,amsthm,amsbsy,dsfont}
\usepackage{feynmp}
\usepackage[force]{feynmp-auto}

\usepackage{a4wide}

\def\tbr#1{{\left\langle #1 \right\rangle}}
\def\sbr#1{{\left[ #1 \right]}}

\def\ri{{\rm i}}

\def\be{\begin{eqnarray}}
\def\ee{\end{eqnarray}}

\usepackage{hyperref}
\hypersetup{
	    hidelinks,
	colorlinks=true,
	allcolors=blue,
	    citecolor=green,
	linktocpage=true,
	pdfhighlight=/N
}
\usepackage{bookmark}

\begin{document}
\begin{fmffile}{graph}

\title{Chiral perturbation theory for GR}
\author{Kirill Krasnov${}^{(1)}$ and Yuri Shtanov${}^{(2,3)}$\\ {}\\
{\it \small ${}^{(1)}$School of Mathematical Sciences, University of Nottingham, NG7 2RD, UK}\\
{\it \small ${}^{(2)}$Bogolyubov Institute for Theoretical Physics, Metrologichna St. 14-b, Kiev 03143, Ukraine}\\
{\it \small ${}^{(3)}$Astronomical Observatory, Taras Shevchenko National University of Kiev,}\\ {\it \small Observatorna St. 3, Kiev 04053, Ukraine}}

\date{\today}
\maketitle

\begin{abstract} We describe a new perturbation theory for General Relativity, with the chiral first-order Einstein--Cartan action as the starting point. Our main result is a new gauge-fixing procedure that eliminates the connection-to-connection propagator. All other known first-order formalisms have this propagator non-zero, which significantly increases the combinatorial complexity of any perturbative calculation. In contrast, in the absence of the connection-to-connection propagator, our formalism leads to an effective description in which only the metric (or tetrad) propagates, there are only cubic and quartic vertices, but some vertex legs are special in that they cannot be connected by the propagator.  The new formalism is the gravity analog of the well-known and powerful chiral description of Yang--Mills theory.
\end{abstract}

\tableofcontents

\section{Introduction}

The action of General Relativity (GR) becomes polynomial in any first-order formalism. Thus, one can tame the algebraic complexity of the perturbative expansion of the Einstein--Hilbert action by introducing an auxiliary connection variable. The tradeoff is that this connection becomes an additional field on top of the metric, with non-vanishing 2-point functions to itself and to the metric. But the number of different vertices one has to consider becomes finite. In the usual metric formulation, the ultimate power of this perturbative first-order formalism is achieved by taking the inverse densitised metric as the basic variable. The action is then cubic in the fields; see, e.g., \cite{Cheung:2017kzx}. 

In the metric/affine-connection formalism, there are propagators of all possible types: metric-metric, metric-connection and connection-connection. Also, these propagators typically are sums of terms with a complicated structure of Lorentz indices, which means that even simple Feynman diagrams generate many terms. One can try to  simplify things by a shift of the connection field that eliminates the metric-connection propagator (see \cite[Eq.~(25)]{Cheung:2017kzx}), but this results in a considerably complicated character of the interaction vertices. At the same time, the connection-connection propagator of this formalism is algebraic, i.e., the connection is a true auxiliary and non-propagating field, present there just to reduce the algebraic complexity of the action. This suggests that there must exist a first-order formalism in which the connection-connection propagator is totally absent. 

The purpose of this paper is to develop such a perturbative formalism for four-dimen\-sio\-nal GR\@. The formalism we describe is based on two elements. First, we use the two-component spinor technique, as is appropriate for computations that are based on the spinor helicity formalism. Second, the formalism we develop is chiral. It relies on special features of four spacetime dimensions and uses self-dual projections. Being chiral means that one of the graviton helicities is described differently from the other. One can anticipate that, at least in four dimensions, the usage of two-component spinor technique (and the associated availability of the powerful Schouten identity) should allow one to write both propagators and vertices more compactly.

It is known that, in four spacetime dimensions, one can impose the Einstein condition by considering only the chiral half of the Levi-Civita connection. The necessary chiral projections are easiest to describe in the spinor notation. In units $8\pi G=1$, the corresponding action (with zero cosmological constant) is 
\begin{equation}\label{action}
	S[\theta,\omega] = 2 \ri \int \Sigma^{AB} \wedge F_{AB} \, ,
\end{equation}
where $A,B=1,2$ are unprimed 2-component spinor indices, the self-dual 2-forms are given by 
\begin{equation}\label{sigma}
\Sigma^{AB} = \frac12 \theta^A{}_{C'} \wedge \theta^{BC'} \, ,
\end{equation}
with $\theta^{AA'}$ being the soldering form (tetrad), and the curvature 2-form $F^{AB}$ is given by
\begin{equation}
	F^{AB} = d \omega^{AB} + \omega^{AC} \wedge \omega_C{}^B \, .
\end{equation}
The object $\omega^{AB}$ is the self-dual part of the spin connection, 
and is locally a one-form with values in the symmetric $\omega^{AB}=\omega^{(AB)}$ second power of the unprimed spinor bundle. Integrating out the connection $\omega^{AB}$ by solving its field equations and substituting the result into the action, one obtains the Einstein--Hilbert action $(1/2) \int R \sqrt{-g}$ for the metric 
\begin{equation}
ds^2 = \theta^A{}_{A'} \theta_A{}^{A'} \, ,
\end{equation}
together with the so-called Holst term (see \cite{Holst:1995pc}) with an imaginary coefficient. The Holst term is a total derivative. So, modulo this imaginary boundary term, the action (\ref{action}) gives an equivalent description of GR\@. This action is obtained by applying the chiral self-dual projection to the first-order Einstein--Cartan action  in terms of the tetrad $\theta^{AA'}$ and the full spin connection (of which $\omega^{AB}$ is the self-dual part). Alternatively, the full non-chiral Einstein--Cartan action can be written as the real part of (\ref{action}). 

The action (\ref{action}) is closely analogous to its, perhaps, better known Yang--Mills (YM) cousin. Thus, modulo (an imaginary) boundary term, the YM action can be written as
\begin{equation}\label{action-YM}
S[A]= -\frac{1}{2g^2} \int \left( F^a_+ \right)^2 \, ,
\end{equation}
where $F^a_+$ is the self-dual projection of the YM field strength, and $a$ is a Lie algebra index. The usual YM action, a multiple of $\int F^2$, can be written as the sum $\int (F_+)^2+(F_-)^2$. One then notes that, in view of the Pontryagin topological term $\int F\wedge F \sim \int (F_+)^2-(F_-)^2$, the integrals of the self-dual (SD) and anti-self-dual (ASD) parts of the field strength are equal to each other, modulo a surface term. This explains why (\ref{action-YM}) is a valid YM action. This action can then be written in the first-order form by introducing a self-dual auxiliary field $B_+$:
\begin{equation}\label{action-YM-first}
S[A,B_+] = \int F^a B_+^a +g^2 \left( B_+^a \right)^2 \, .
\end{equation}
Integrating out $B_+^a$, one gets back (\ref{action-YM}). The action (\ref{action-YM-first}) has only a simple cubic vertex. Moreover, there is a natural gauge-fixing procedure that leads to the absence of the propagator of the auxiliary field $B_+$ with itself, see, e.g., Section~2 of \cite{Krasnov:2016emc} for details. 

There exists an effective formalism for this YM perturbation theory.
In this formalism, one of the legs in the cubic vertex becomes special in that two such legs should never join to form a propagator. Feynman graphs in which such connections were present would exactly cancel the contributions from the quartic vertex of the usual YM perturbation theory, see, e.g., \cite[Eq.~(55)]{Cofano:2015jva} for more details.  
So, in effect, the main advantage of the formalism (\ref{action-YM-first}) is that it puts the YM cubic vertex in the form that eliminates the need to consider the quartic vertex. 

It is easy to pass from the chiral description of YM provided by (\ref{action-YM-first}) to the self-dual YM theory, in which the only solutions are YM field configurations with vanishing self-dual part of the field strength. This is achieved simply by setting $g=0$ in the above action. This eliminates the gauge field to gauge field propagator, leaving only the 2-point function between $B_+$ and $A$ non-zero. The diagrams that can be constructed in the self-dual YM are a subset of the diagrams of the full YM, which means that a subset of the full YM amplitudes is correctly captured by the self-dual theory, see \cite{Krasnov:2016emc} for more on this.

Our gravity formalism based on (\ref{action}) is precisely analogous to the above description. Firstly, as we already noted, a gauge is available in which the auxiliary field $\omega^{AB}$ has a vanishing 2-point function with itself. This is by no means trivial, and relies on a delicate matching between the number of components in the perturbation of the tetrad, in the connection, and in the Lagrange multipliers added in the process of gauge fixing. We will explain the main idea of our gauge-fixing procedure below. 

Secondly, there are only cubic and quartic vertices, as is clear from expanding (\ref{action}) perturbatively. The arising propagators will be of two types: tetrad-tetrad and tetrad-connection. The latter has a factor of momentum in the numerator, while the former is the standard $1/k^2$ times a product of the spinor metrics. One then notes that the derivative present in the tetrad-connection propagator can be assigned to the connection legs of the vertices. This leads to an effective formalism in which only the tetrad propagator is present, and all vertices contain two derivatives as is appropriate for a gravity theory. However, one needs to mark some legs of vertices (those that come from the connection $\omega$) as special, and to respect the rule that special legs cannot connect to each other. Feynman graphs in which such connections were present would result in contact terms that would need to be cancelled by higher order vertices. Thus, as in the case of chiral YM, the price to pay for low valency of the vertices is that some vertex legs are marked as special. 

Finally, there is also a way to pass from (\ref{action}) to a version of the theory where only (anti-)\,self-dual configurations are solutions. This is achieved by removing the $\omega\omega$ term of the curvature $F$ from the action, which eliminates the tetrad-to-tetrad propagator. The corresponding action for self-dual GR is contained in the bosonic part of the action in Section 6 of \cite{Siegel:1992wd}. 

As we already noted, our main result is a new gauge-fixing procedure that eliminates the connection-connection propagator. It will be useful to describe this gauge-fixing procedure qualitatively already in the Introduction. 

\bigskip
\noindent {\bf Gauge-fixing procedure.} When expanded perturbatively around the Minkowski background, the action (\ref{action}) is a function of the tetrad perturbation $h^{AA'BB'}$ with its 16 components, as well as the chiral spin connection $\omega^{ABCC'}$, which is symmetric in its first two indices, and so has $3\times 4=12$ components. However, one quickly finds that the free Lagrangian (i.e., the collection of terms that are quadratic in the perturbations) is independent of the component $h^{A(A'}{}_{A}{}^{B')}$ of the tetrad perturbation. This component transforms non-trivially and irreducibly with respect to the anti-self-dual chiral half of the Lorentz group, and the gauge in which it is set to zero in the complete action is the most natural one. 

Setting $h^{A(A'}{}_{A}{}^{B')}$ to zero results in 13 components in the tetrad perturbation, plus 12 components in the connection. The first step of our gauge-fixing procedure is to add 4 more fields $\lambda^{AA'}$ to the linearised action with the purpose of fixing the diffeomorphism gauge freedom. The most economic way of doing this is to absorb these new fields into the connection field $\omega^{ABCC'}$ as its extra components. This is done by adding to the connection, originally symmetric in $\scriptstyle AB$, a new, $\scriptstyle AB$ anti-symmetric part, which contains precisely 4 new fields. We show that this can be done in such a way as to produce a de Donder type gauge-fixing term in the Feynman form, schematically $\lambda \partial h + \lambda^2$. Overall, the diffeomorphism gauge freedom is gauge-fixed by extending the connection to contain 4 more components. The new connection field is again $\omega^{ABCC'}$, but without any symmetry property.

It remains to fix the gauge freedom of the self-dual chiral part of the Lorentz group. A straightforward way to do it would be to set to zero the component $h^{(A}{}_{A'}{}^{B)A'}$ of the tetrad perturbation. One would then deal with 10 components of the metric perturbation plus 16 components of the connection. However, because of the mismatch between these numbers, there would be a non-trivial connection-connection propagator, as in all previous versions of the gravitational perturbation theory.

Our main new finding is a different way to fix the self-dual chiral half of Lorentz gauge freedom. It is done by using a Lorentz-type gauge condition  $\partial^{CC'}\omega^{AB}{}_{CC'} = 0$ with a derivative on the connection rather than an algebraic condition on the tetrad. This gauge condition requires introduction of 3 new Lagrange multipliers forming a symmetric spinor $\lambda^{AB}$. Together with the 13 components already contained in $h^{AA'BB'}$, this gives 16 field components, which is the same number as in the connection field (extended by $\lambda^{AA'}$). Matching the number of components in the (extended) tetrad and connection fields is necessary to get rid of the connection-connection propagator.  

In our gauge-fixing procedure, the space of fields $h^{AA'BB'}$,  $\omega^{ABCC'}$, $\lambda^{AA'}$, $\lambda^{AB}$ is separated into two sets of conjugate pairs that decouple in the free Lagrangian and greatly facilitates the procedure of inverting the kinetic term. This separation is non-trivial and is described in the main text, but its essential elements are as follows. The part $h^{(AB)(A'B')}$ of the tetrad perturbation, symmetric with respect to both pairs of indices, is combined together with $\lambda^{AB} \epsilon^{A'B'}$, where $\epsilon^{A'B'}$ is the spinor metric, into a new field $H^{ABA'B'} = H^{(AB)A'B'}$, which is no longer symmetric in the primed spinor indices, but remains symmetric in the unprimed ones. There is then a certain combination of $\omega^{ABCC'}$ and $\lambda^{AA'}$, which we call $\Omega^{ABCC'} = \Omega^{(AB)CC'}$ and which is symmetric in its first two indices.  In the free Lagrangian it couples to the field $H^{ABA'B'}$, both fields having 12 components. The kinetic term for this pair of fields is trivial to invert, as its derivative part is just the chiral Dirac operator, see (\ref{Lkin}). 

In addition to the pair of fields $(H^{ABA'B'},\, \Omega^{ABCC'})$, there is another pair. The trace part $h^{AA'}{}_{AA'}$ of the tetrad perturbation combines with $h^{(AB)A'}{}_{A'}$ into a four-component field $h^{AB}$ without any symmetry property. In addition, there are four remaining components of the connection, a field we call $\omega^{AA'}$, and the derivative part of the kinetic term for this pair of fields is again a chiral Dirac operator, see the second term in (\ref{Lkin}).

All in all, our gauge-fixing procedure resulted in matching the number of (extended) tetrad and connection fields, and separation of the free part of the theory into two decoupled sectors. This leads us to very simple propagators, which is the main result of our analysis. 

\bigskip

We work in the signature mostly plus. The organisation of the rest of the paper is as follows. In Sec.~\ref{sec:expansion}, we describe the expansion of the chiral action in basic fields and introduce the necessary notation. In Sec.~\ref{sec:gauge}, which is the central section of this paper, we describe our non-trivial procedure of gauge fixing, obtain the field variables that diagonalise and simplify the free Lagrangian, and calculate the propagators. In Sec.~\ref{sec:interaction}, we discuss the structure of interaction and introduce the formalism of effective vertices with marked lines.  As a test of our formalism, we apply it to the calculation of three-point amplitudes in Sec.~\ref{sec:amplitudes}, some simple off-shell currents in Sec.~\ref{sec:currents}, and 2-to-2 graviton scattering amplitudes in Sec.~\ref{sec:scattering}.  We formulate our conclusions in Sec.~\ref{sec:conclusions}. Appendix describes an alternative procedure to arrive at our gauge-fixing. 

\section{Expansion of the action}
\label{sec:expansion}

We expand the fields around the Minkowski space configuration. Since the background connection is zero, $\omega^{AB}$ describes the total connection. From now on we denote by $\theta^{AA'}$ the Minkowski spacetime tetrad, and its perturbation is denoted as $h^{AA'}$. We then split action (\ref{action}) into the free and interaction parts:
\begin{align}\label{free}
	S_0 &= \ri \int \theta^A{}_{A'} \wedge \theta^{BA'} \wedge \omega_A{}^C \wedge \omega_{CB} + \ri \int \left( \theta^A{}_{A'} \wedge h^{BA'} + h^A{}_{A'} \wedge \theta^{BA'} \right) \wedge d \omega_{AB} \, , \\
	S_{\rm int} &= 2 \ri \int \theta^A{}_{A'} \wedge h^{BA'} \wedge \omega_A{}^C \wedge \omega_{CB} + \ri \int h^A{}_{A'} \wedge h^{BA'} \wedge \left( d \omega_{AB} + \omega_A{}^C \wedge \omega_{CB} \right) \, . \label{inter}
\end{align}

We expand all forms in the basis of background 1-forms $\theta^{AA'}$. Thus, we have
\begin{equation}\label{tetex1}
	h^{AA'} = h^{AA'}{}_{MM'} \theta^{MM'} \, , \qquad \omega^{AB} = \omega^{AB}{}_{MM'} \theta^{MM'} \, , 
\end{equation}
as well as
\begin{equation}\label{tetex2}
	d\omega^{AB} = \partial_{MM'} \omega^{AB}{}_{NN'} \theta^{MM'}\wedge \theta^{NN'} .
\end{equation}

Under diffeomorphisms generated by an infinitesimal vector field $\xi^\mu = \xi^{AA'} \theta_{AA'}^\mu$, and local SL(2,C) gauge transformations with the infinitesimal parameters $\phi^{AB},\bar{\phi}^{A'B'}$, our variables transform as
\begin{align}
	\delta_\xi h^{AA'}{}_{BB'} &= \partial_{BB'} \xi^{AA'} \, , &\delta_\xi \omega^{AB}{}_{CC'} &= 0 \, , \label{xitrans} \\
	\delta_\phi h^{AA'}{}_{BB'} &= {} - \phi^A{}_B \epsilon^{A'}{}_{B'} - \bar \phi^{A'}{}_{B'} \epsilon^A{}_B \, , 
	&\delta_\phi \omega^{AB}{}_{CC'} &= \partial_{CC'} \phi^{AB} \, . \label{phitrans}
\end{align}

\subsection{Parametrisation of the tetrad perturbation}

To exhibit the structures arising, let us decompose the tetrad perturbation into its irreducible components
\begin{equation}
	h^{AA'BB'} = h^{(AB)(A'B')} + h^{(AB)}\epsilon^{A'B'} + h^{(A'B')}\epsilon^{AB} + h \epsilon^{AB}\epsilon^{A'B'},
\end{equation}
which also defines all the components on the right hand-side. In fact, it will prove to be convenient to combine the second and fourth terms here, and define a new field $h^{AB}=h^{(AB)}+h\epsilon^{AB}$ that is no longer $\scriptstyle AB$ symmetric. Thus, let us instead use the following parametrisation
\begin{equation}
	h^{AA'BB'} = h^{ABA'B'} + h^{AB}\epsilon^{A'B'} + h^{A'B'}\epsilon^{AB} .
\end{equation}
It should now be kept in mind that $h^{ABA'B'}$ is symmetric in its both pairs of indices, $h^{AB}$ does not have any symmetry, and $h^{A'B'}$ is symmetric.\footnote{To minimize the notation, we thus distinguish between $h^{AA'BB'}$ and its complete symmetrisation $h^{ABA'B'} = h^{(AB)(A'B')}$ by the position of indices.} 

\subsection{Imposing a partial gauge}

Using the action of one chiral half of the Lorentz group [see the first equation in (\ref{phitrans})], we can eliminate the $h_{A'B'}$ part of the tetrad perturbation. This part does not enter into the linearised action, and only appears in the interaction terms. We will simplify things considerably by killing this part from the start. With this part set to zero, we have the following parametrisation of the tetrad perturbation:
\begin{equation}\label{hdeco}
	h^{AA'BB'} = h^{ABA'B'} + h^{AB}\epsilon^{A'B'} .
\end{equation}

\section{Gauge fixing}
\label{sec:gauge}

Our strategy of fixing the diffeomorphism gauge freedom will consist in removing the symmetry of $\omega^{AB}$ and modifying the action in an appropriate way [the antisymmetric part of $\omega^{AB}$ would drop from the unmodified action (\ref{free}), (\ref{inter})]. The net result of all this will be generation of reasonable gauge-fixing terms for the original action. After that, we fix the other half of the Lorentz gauge freedom by adding new Lagrange multipliers. The whole procedure will greatly simplify the free part of the action, splitting it into two autonomous sectors, and eliminating the connection-connection propagator, as we will see below. 

The unmodified action is calculated by expanding the forms according to (\ref{tetex1}), (\ref{tetex2}) and using the arising orientation form to contract the spinor indices in (\ref{free}) and (\ref{inter}).  Specifically, one uses the relation
\begin{equation}
\theta^{MM'}\wedge \theta^{NN'} \wedge \theta^{RR'}\wedge \theta^{SS'}=  \epsilon^{MM'NN'RR'SS'} \upsilon \, ,
\end{equation}
where $\upsilon$ is the space-time volume form, and $\epsilon^{MM'NN'RR'SS'}$ is the internal-space orientation.  This last has two known representations, each consisting of two mutually complex-conjugate terms, which we denote appropriately (omitting indices)\footnote{Our choice of orientation differs by sign from the usual one.}:
\begin{align}
\text{Penrose representation:} \quad &\epsilon = \epsilon_{\rm P} + \bar \epsilon_{\rm P} \, , \quad &\epsilon_{\rm P}^{MM'NN'RR'SS'} = \ri\, \epsilon^{MS} \epsilon^{NR} \epsilon^{M'R'}  \epsilon^{N'S'}  \, , \label{orP} \\	
\text{Wald representation:} \quad &\epsilon = \epsilon_{\rm W} + \bar \epsilon_{\rm W} \, , \quad &\epsilon_{\rm W}^{MM'NN'RR'SS'} = \ri\, \epsilon^{MR} \epsilon^{NS} \epsilon^{M'N'}  \epsilon^{R'S'} \, . \label{orW}
\end{align}

Our idea of modifying the action is to perform contraction of indices by using separately two parts of orientation in (\ref{orP}) or (\ref{orW}) with different position of indices of $\omega^{AB}$ in the action for each part (and treating now $\omega^{AB}$ as a connection without any symmetry).  We proceed to describing this in detail.\footnote{The third possible splitting $\epsilon^{MM'NN'RR'SS'} = \ri\, \epsilon^{MS} \epsilon^{NR} \epsilon^{M'N'} \epsilon^{R'S'}  + \text{c.c.}$ does not lead to anything interesting.}

\subsection{Kinetic term}

Consider first the kinetic term in (\ref{free}).  With respect to this term, we use the following schematic procedure\footnote{This procedure is not unique, e.g., one could transpose the indices of $\omega^{AB}$ in the first term of (\ref{kinsplit}) rather than in the second one.  This, however, will lead to a similar final result.}:
\begin{equation} \label{kinsplit}
	\left( \theta^A{}_{A'} \wedge h^{BA'} \wedge d \omega_{AB} \right)_{\textstyle \epsilon_{\rm P}} + \left( \theta^A{}_{A'} \wedge h^{BA'} \wedge d \omega\underline{_{BA}} \right)_{\textstyle \bar \epsilon_{\rm P}} \, ,
\end{equation}
where the subscripts $\epsilon_{\rm P}$ and $\bar \epsilon_{\rm P}$ mean contraction with the corresponding parts in (\ref{orP}), and we have underlined the transposed indices in $\omega^{AB}$.  This is equivalent to saying that the term with the symmetric part $\omega^{(AB)}$ of the connection is contracted with the complete orientation spinor $\epsilon_{\rm P} + \bar \epsilon_{\rm P} = \epsilon$, while the term containing the antisymmetric part $\omega^{[AB]}$ is contracted with $ \epsilon_{\rm P} - \bar \epsilon_{\rm P}$. The result is
\begin{equation}\label{kin0}
	L_{\rm kin} = 2 h^{AA'BB'} \left( \partial_{BA'} \omega^C{}_{ACB'} - \partial_{CB'} \omega_A{}^C{}_{BA'} \right) \, .
\end{equation}

The presence of the antisymmetric part in the connection, $\omega^{[AB]CA'} = \epsilon^{AB} \lambda^{CA'}$, adds to the original action an extra term
\begin{equation}\label{lam}
	2 \lambda^{AA'} \partial_{BB'} \left( h_A{}^{B'B}{}_{A'} + h^B{}_{A'A}{}^{B'} \right) \, ,
\end{equation}
which can be seen to be precisely the combination that leads to the de~Donder gauge for the tetrad perturbation. 

Using decomposition (\ref{hdeco}) for the tetrad perturbation, from (\ref{kin0}) we obtain
\begin{align}\label{Lkin0}
	L_{\rm kin} &= 2 h^{ABA'B'} \left( \partial_{BA'}\omega^C{}_{ACB'} - \partial_{CA'} \omega_A{}^C{}_{BB'} \right) \nonumber \\ &\quad {} + 2 h^{AB} \left( \partial_{BA'} \omega^C{}_{AC}{}^{A'} + \partial_{CA'} \omega_A{}^C{}_B{}^{A'} \right) \, . 
\end{align}
This expression can be written identically as
\begin{equation} \label{Lkin1}
L_{\rm kin} = - 2 \left[ h^{ABA'B'} - h^{(AB)} \epsilon^{A'B'} \right] \partial_{CA'} \left( \omega_A{}^C{}_{BB'} + \epsilon^C{}_B \omega^D{}_{ADB'} \right) + 4 h^{AB} \partial_{BB'} \omega^C{}_{AC}{}^{B'}\, .
\end{equation}
We note the appearance of two special combinations of the connection, coupling to the tetrad perturbations, for which we introduce the notation
\begin{align}\label{om}
\omega^{AA'} &= \omega^{CA}{}_C{}^{A'} \, , \\
\Omega^{ABCA'} &= \omega^{ACBA'} + \epsilon^{CB} \omega^{AA'} = \omega^{ABCA'} - \epsilon^{BC} \omega^D{}_D{}^{AA'} \, . \label{Om}
\end{align}
In deriving the last equality, we have used the identity
\begin{equation}
	\omega^{C}{}_{CAA'} = \omega^C{}_{ACA'} - \omega_A{}^C{}_{CA'} \, .
\end{equation}
Spinor (\ref{Om}) is symmetric in its first two indices by construction, hence, has 12 independent components, while (\ref{om}) has 4 components. Together, these fields completely describe the connection with its 16 components. Note that the first relation in (\ref{Om}) already expresses the connection field $\omega^{ABCA'}$ in terms of new variables:
\begin{equation}\label{w}
	\omega^{ABCA'} = \Omega^{ACBA'} + \epsilon^{CB} \omega^{AA'} \, .
\end{equation} 

The conjugate fields in (\ref{Lkin1}) are $h^{ABA'B'}$ with 9 components, and $h^{AB}$ with 4 components. It is clear then that the kinetic term is still degenerate. On the other hand, only one of the two chiral halves of the Lorentz gauge freedom has been fixed thus far. To fix the other half, we add additional Lagrange multipliers, to impose a version of the Lorentz gauge. Specifically, we add the term $\lambda^{AB} \epsilon^{M'N'}$ to the expression in the first bracket of (\ref{Lkin1}). We  introduce a new name for the combination that arises in this way: 
\begin{equation}\label{H}
	H^{ABA'B'}:= h^{ABA'B'} + \left[\lambda^{AB} - h^{(AB)} \right] \epsilon^{A'B'}\, .
\end{equation}
This adds additional 3 components to the metric perturbation fields. Given that the field $\lambda^{AB}$ is independent, this procedure makes the fields $H^{ABA'B'}$ and $h^{AB}$ completely independent. The new spinor (\ref{H}) is symmetric in its first two indices, thus having 12 independent components, and is conjugate to $\Omega^{ABCA'}$.  Using (\ref{hdeco}) and (\ref{H}), we can express the tetrad perturbation and the new Lagrange multiplier in terms of new variables:
\begin{align}\label{hdeco1}
	h^{AA'BB'} &= H^{AB(A'B')} + h^{AB} \epsilon^{A'B'} \, , \\
	\lambda^{AB} &= h^{(AB)} - \frac12 H^{ABC'}{}_{C'} \, . \label{ldeco}
\end{align}

This completely gauge fixes the kinetic term, so that the kinetic part of the Lagrangian reads
\begin{equation}\label{Lkin}
	L_{\rm kin} = - 2H^{ABA'B'}  \partial_{CA'} \Omega_{AB}{}^C{}_{B'} + 4h^{AB}\partial_{BA'} \omega_{A}{}^{A'} \, .
\end{equation}
It is decoupled into two sectors $\left( H , \Omega \right)$ and $\left( h ,  \omega \right)$.

\subsection{Potential term}

It turns out that the potential term can be diagonalised in new connection components (\ref{om}) and (\ref{Om}) by using the splitting of the orientation in the Wald form.  Specifically, similarly to (\ref{kinsplit}), we modify the first term in (\ref{free}) as
\begin{equation}
	\left( \theta^A{}_{A'} \wedge \theta^{BA'} \wedge \omega_A{}^C \wedge \omega_{CB} \right)_{\textstyle \epsilon_{\rm W}} + \left( \theta^A{}_{A'} \wedge \theta^{BA'} \wedge \omega_A{}^C \wedge \omega\underline{_{BC}} \right)_{\textstyle \bar \epsilon_{\rm W}} \, ,
\end{equation}
in which, again, we underlined the transposed indices in $\omega^{AB}$. The result is 
\begin{equation}\label{Lpot}
	L_{\rm pot} = - \omega^{ABCA'} \omega_{ABCA'} + 2 \omega^{AB}{}_B{}^{A'} \omega^C{}_{ACA'} = - \Omega^{ABCA'} \Omega_{ABCA'} + 2 \omega^{AA'} \omega_{AA'} \, .
\end{equation}
Here, we have made the substitution (\ref{w}) to obtain the last equality.  Remarkably, the potential term also decouples in two sectors, which makes the process of finding propagators trivial; they can be read off from the gauge-fixed linearised action.

The antisymmetric part $\omega^{[AB]CA'} = \epsilon^{AB} \lambda^{CA'}$ in (\ref{Lpot}) adds to the original action another term, quadratic in the Lagrange multiplier $\lambda^{AA'}$, which, together with (\ref{lam}) and insertion of $\lambda^{AB}$ into (\ref{Lkin1}), fixes the diffeomorphism gauge in the Feynman form as follows:
\begin{align}
L_{\rm g.f.} &=	2 \lambda^{AA'}\left[ \partial_{BB'} \left( h_A{}^{B'B}{}_{A'} + h^B{}_{A'A}{}^{B'} \right) - 2 \partial_{BA'} \lambda_A{}^B \right] - 4 \lambda^{AA'} \lambda_{AA'} \nonumber \\ &=  4 \lambda^{AA'} \left( \partial^{BB'} H_{ABA'B'} - \partial^B{}_{A'} H_{AB}{}^{C'}{}_{C'} + \partial^B{}_{A'} h_{AB} \right) - 4 \lambda^{AA'} \lambda_{AA'} \, .
\end{align}

\subsection{Propagators}

Now that the $(H,\Omega)$ and $(h,\omega)$ sectors are decoupled, it is easy to derive the propagators. To this end, we couple all fields to currents. The complete free Lagrangian with currents is
\begin{align}\label{Lnew}
	L_{\rm free} = {} -  \Omega^{ABCA'} \Omega_{ABCA'} - 2\Omega_{ABCA'} \partial^C{}_{B'} H^{ABB'A'} + 2\omega^{AA'} \omega_{AA'} + 4 \omega_{AA'} \partial_B{}^{A'} h^{AB} \nonumber \\ {} + J_{ABCA'} \Omega^{ABCA'} + J_{AA'} \omega^{AA'} + J_{ABA'B'} H^{ABA'B'} + J_{AB} h^{AB} \, , 
\end{align}
where the currents for our new fields have respective symmetries in indices.

From Lagrangian (\ref{Lnew}), we obtain the field equations
\begin{align} 
	\frac{\delta}{\delta \Omega} : \quad &\Omega^{ABCA'} = \frac12 J^{ABCA'} - \partial^C{}_{B'} H^{ABB'A'} \, , \label{eq-Om} \\ 
	\frac{\delta}{\delta \omega} : \quad  &\omega^{AA'} = \partial^{BA'} h^A{}_B - \frac14 J^{AA'} \, , \label{eq-Phi} \\
	\frac{\delta}{\delta H} : \quad &\partial^C{}_{A'} \Omega_{ABCB'} + \frac12 J_{ABA'B'} = 0 \, ,  \label{eq-H} \\
	\frac{\delta}{\delta h} : \quad &\partial_{BA'} \omega_A{}^{A'} + \frac14 J_{AB} = 0 \, .  \label{eq-h}
\end{align}

Substituting (\ref{eq-Om}) and (\ref{eq-Phi}) into (\ref{eq-H}) and (\ref{eq-h}), we obtain
\begin{align}
	&\partial^C{}_{A'} J_{ABCB'} +  \Box H_{ABA'B'} + J_{ABA'B'} = 0 \, , \\
	&2\Box h_{AB} + \partial_{BA'} J_A{}^{A'} - J_{AB} = 0 \, ,
\end{align}	
where $\Box = \partial^A{}_{A'} \partial_A{}^{A'}$. From these and (\ref{eq-Om}), (\ref{eq-Phi}), we get
\begin{align}
	\Box H_{ABA'B'} &= - \partial^C{}_{A'} J_{ABCB'} - J_{ABA'B'} \, , \label{eq1} \\
	\Box h_{AB} &= \frac12 \left( J_{AB} - \partial_{BA'} J_A{}^{A'} \right)\, , \\
	\Box \Omega_{ABCA'} &=  \partial_{CC'} J_{AB}{}^{C'}{}_{A'} \, , \\
	\Box \omega^{AA'} &= \frac12 \partial^{BA'} J^A{}_B \, . \label{eq4}
\end{align}

The generating Lagrangian can be calculated as
\begin{equation}
	L_W  = \frac12 \left( J_{ABCA'} \Omega^{ABCA'}  + J_{AA'} \omega^{AA'} + J_{ABA'B'} H^{ABA'B'} + J_{AB} h^{AB} \right) \, ,
\end{equation}
where we need to substitute the solutions for the fields in terms of the sources.  Using (\ref{eq1})--(\ref{eq4}), we obtain the result:
\begin{align}\label{propag}
	L_W &=  - \frac12 J_{ABA'B'} \Box^{-1} J^{ABA'B'} + \frac14 J_{AB} \Box^{-1} J^{AB} \nonumber \\ &\quad {} +  J^{ABCA'} \Box^{-1} \partial_{CB'} J_{AB}{}^{B'}{}_{A'} - \frac12 J^{AA'} \Box^{-1} \partial_{BA'} J_A{}^B \, .
\end{align}

All propagators can be read off directly from here. We will only need the $HH$ and $H\Omega$ propagators for what follows. The $HH$ propagator is given by
\begin{equation}\label{prop-HH}
\left \langle H_{ABA'B'}(k) H^{MNM'N'}(-k) \right \rangle = \frac{1}{\ri k^2}\, \epsilon_A{}^{(M} \epsilon_B{}^{N)} \epsilon_{A'}{}^{M'} \epsilon_{B'}{}^{N'} \, , \qquad
\begin{fmfgraph*}(70,20)
\fmfleft{i1}
\fmfright{i2}
\fmf{dbl_plain}{i1,i2}
\end{fmfgraph*}
\end{equation}
which we depict by a double straight line. The symmetrisation of the propagator in the unprimed spinor indices is necessary because the field $H^{ABA'B'}$ is $\scriptstyle AB$ symmetric. The $\Omega H$ propagator is given by 
\begin{equation}\label{prop-OH}
\left \langle \Omega_{ABCC'}(k) H^{MNM'N'}(-k) \right \rangle = \frac{1}{k^2}  \, \epsilon_A{}^{(M}\epsilon_B{}^{N)}  k_C{}^{M'} \epsilon_{C'}{}^{N'}\, , \qquad
\begin{fmfgraph*}(70,20)
\fmfleft{i1}
\fmfright{i2}
\fmf{dbl_wiggly}{i1,v}
\fmf{dbl_plain}{v,i2}
\end{fmfgraph*}
\end{equation}

We note that we can understand the $\Omega H$ propagator as the result of applying a derivative to the $HH$ propagator. Indeed, we see from (\ref{eq-Om}) that, in the absence of the current, 
\be\label{Omega-H*}
\Omega_{ABCC'} =  - \partial_{CA'} H_{AB}{}^{A'}{}_{C'}.
\ee
This relation also holds for the propagators. Indeed, we see that
\be\label{prop-relation}
\left\langle \Omega_{ABCC'}(k) H^{MNM'N'}(-k) \right\rangle = - \ri k_{CA'} \left \langle H_{AB}{}^{A'}{}_{C'}(k) H^{MNM'N'}(-k) \right \rangle \, .
\ee
As we shall see below, this means that, for all practical purposes, we can replace $\Omega$ with its expression (\ref{Omega-H*}), keeping in mind, however, that the corresponding copy of $H$ is related to the connection and is special. This will be discussed in more details below.

\section{Interaction}
\label{sec:interaction}

\subsection{The $h h \partial \omega$ term}

For this term, our orientation-splitting procedure gives little simplification.  Using the procedure similar to   (\ref{kinsplit}), we obtain
\begin{equation} \label{hhdw0}
	\ri h^A{}_{A'} \wedge h^{BA'} \wedge d \omega_{AB} \quad \to \quad 2 h^{AA'BB'} h^C{}_{A'}{}^{DC'}  \partial_{DB'} \omega_{ACBC'} \, .
\end{equation}

In this term, we first can replace $\omega^{ABCA'}$ by $\Omega^{ABCA'}$.  Indeed, the difference between them, according to (\ref{Om}) is $\epsilon^{BC} \omega^D{}_D{}^{AA'}$, which is our Lagrange multiplier for fixing the diffeomorphism gauge.  With decomposition (\ref{hdeco1}) taken into account, this component then becomes
\begin{align}\label{L1}
	L_1 = - 2H^{AC(B'C')} H^{BD}{}_{(A'C')} \partial_{DB'} \Omega_{ABC}{}^{A'} - 2h^{AB} h^{CD} \partial_{DA'} \Omega_{ACB}{}^{A'} \nonumber \\ {} + 2h^{AB} H_{CB(A'B')} \partial_D{}^{B'} \Omega_A{}^{CDA'} \, .
\end{align}
Note that the connection component namely $\omega_{AA'}$ does not appear in this part of interaction. It will only enter the parts $L_2$ and $L_3$ of the combined type $(h+hh)\omega\omega$. 

It would be nice to get rid of the symmetrization in (\ref{L1}).  We note that [see (\ref{H})]
\begin{equation}
	H^{AB(A'B')} = H^{ABA'B'} + \left[ h^{(AB)} - \lambda^{AB} \right] \epsilon^{A'B'} \, .
\end{equation} 
Hence, we could shift the $H$'s in (\ref{L1}) by the Lagrange multiplier as follows:
\begin{equation}\label{H-lambda-shift}
	H^{AB(A'B')} \to H^{AB(A'B')} + \lambda^{AB} \bar \epsilon^{A'B'} = H^{ABA'B'} + h^{(AB)} \bar \epsilon^{A'B'} \, .
\end{equation}
The problem is that this shift will modify the {\em Lorentz\/} constraint by the presence of $\lambda^{AB}$ in it.  In other words, there will be new non-trivial terms quadratic in the added Lagrange multipliers, which may be problematic.

We postpone the full analysis of possible non-linear gauges to a separate publication, where we plan to return to this problem using the BRST formalism. The only cubic vertex that is important for our later purposes will be the $HH\partial \Omega$ term, which we will be written below, and for which the presence of symmetrization in (\ref{L1}) will be of no importance.

\subsection{Other components}

In deriving the other components of interaction we might also use the procedure of splitting the orientation, which we used to deal with the free term.  It turns out, however, that they do not simplify these expressions. Hence, we compute the remaining components of the interaction directly using (\ref{inter}):
\begin{align} 
L_2 &= 2h^{AA'BB'} \left(\omega^C{}_{ABA'} \omega^D{}_{CDB'} - \omega^C{}_{ADB'} \omega^D{}_{CBA'} \right) \, ,  \\ L_3 &= h^{AA'BB'} h^C{}_{A'}{}^{DC'}  \left( \omega_A{}^E{}_{DB'} \omega_{ECBC'} - \omega_A{}^E{}_{BC'} \omega_{ECDB'} \right) \, . 
\end{align}	
Both these parts do not depend on the antisymmetric part $\omega^{[AB]CA'}$ of the connection. 

We would like to use non-linear gauge-fixing procedure in which the Lagrange multipliers added to gauge-fix the kinetic term are also added at the level of the interaction terms. We would also like to work with the fields $H^{ABA'B'}$, $h^{AB}$, $\Omega^{ABCA'}$, and $\omega^{AA'}$, for which the propagators are known. 

If we are to avoid terms quadratic in the Lagrange multiplier $\lambda^{AA'}$ that also depend on the fields, we can only add terms linear in $\lambda^{AA'} = \omega^B{}_B{}^{AA'}$ to modify the diffeomorphism constraint by another non-linear term. The option with the simplest such result is obtained if we substitute $\omega^{ABCA'} \to \Omega^{ABCA'}$ for the (1,4) instances of $\omega$ in the products $\omega_1 \omega_2 - \omega_3 \omega_4$, and leave the other two $\omega$'s intact. These components then read
\begin{align}\label{L2}
	L_2 &= 2h^{AA'BB'} \Omega^{CD}{}_{AB'} \Omega_{CDBA'} - 4 h^{AA'BB'} \omega^C{}_{B'}  \Omega_{ACBA'} \, , \\
		L_3 &=  \omega^{AA'} \left(\Omega_{AC}{}^D{}_{C'} h^{CB'BC'} h_{BB'DA'} + \Omega_{BC}{}^D{}_{C'} h_A{}^{B'BC'} h^C{}_{B'DA'} \right) \nonumber \\ &\quad {} +  h^{AA'BB'} h^C{}_{A'}{}^{DC'} \left( \Omega_A{}^F{}_{DB'}  \Omega_{BFCC'} - \Omega_{AB}{}^F{}_{C'}  \Omega_{CFDB'} \right) 	\, . \label{L3}
\end{align}
Again, we can replace the $h^{AA'BB'}$ field here with its expression in terms of $H^{ABA'B'}$ and $h^{AB}$ fields. For our later purposes we will only need the $H\Omega\Omega$ vertex.

\subsection{Effective $HHH$ vertex}

As we shall see in what follows, the most important role in the tree level computations is played by the $HH\partial \Omega$ vertex given by

\begin{equation}\label{main-vertex-prelim}
    -2\ri H^{AS}{}_{M'}{}^{R'} H^{BR}{}^{M'S'} \partial_{RR'} \Omega_{ABSS'}
   \qquad 
\begin{gathered}
\begin{fmfgraph*}(100,100)
\fmftop{i1,i2}
\fmfbottom{i3}
\fmf{dbl_plain}{i1,v1}
\fmf{dbl_plain}{i2,v1}
\fmf{dbl_wiggly}{i3,v1}
\fmflabel{1}{i1}
\fmflabel{2}{i2}
\fmflabel{3}{i3}
\end{fmfgraph*}
\end{gathered}
\end{equation}

\bigskip
\noindent where the factor of the imaginary unit in front is the one in the exponent of $e^{\ri S}$ in the path integral. When the $\Omega$ leg of this vertex is internal, one has to insert the $\Omega H$ propagator (\ref{prop-OH}) into this leg. However, this propagator can be obtained by applying a derivative to the $HH$ propagator, as we discussed in (\ref{prop-relation}). When this leg is external, one has to substitute into it the corresponding state, which is again obtained (\ref{omega-state}) by applying the derivative to the $H$ state. This means that we can always replace $\Omega$ by its expression (\ref{Omega-H*}), in particular in this vertex, and work with an effective $HHH$ vertex. The only subtlety is that one has to keep track of what used to be the $\Omega$ leg, and remember that there is no $\Omega\Omega$ propagator, which means that two $\Omega$ legs can never contract. This can be taken care of by putting a cilia next to the corresponding leg. The cilia is indicated by a bullet symbol, which we put both as an index of the corresponding copy of $H$, as well as as a label on the corresponding leg of the vertex.

With these conventions, the new effective $HHH$ vertex reads
\fmfcmd{%
style_def dotted expr p =
draw_double p;
filldraw fullcircle scaled 5 shifted point length(p)/4 of p
enddef;}
\begin{equation}\label{main-vertex}
2\ri H^{AS}{}_{M'}{}^{R'} H^{BR}{}^{M'S'} \partial_{RR'} \partial_{SK'} H^\bullet_{AB}{}^{K'}{}_{S'}  \qquad 
\begin{gathered}
\begin{fmfgraph*}(100,100)
\fmftop{i1,i2}
\fmfbottom{i3}
\fmf{dbl_plain}{i1,v1}
\fmf{dbl_plain}{i2,v1}
\fmf{dotted}{v1,i3}
\fmflabel{1}{i1}
\fmflabel{2}{i2}
\fmflabel{3}{i3}
\end{fmfgraph*}
\end{gathered}
\end{equation}
It is second-order in derivatives, as is appropriate for a gravity theory. 

\subsection{Another effective vertex}

The other vertex that is relevant for our tree level computations is $H\Omega\Omega$. This  can be read-off from (\ref{L2}) and reads
\begin{equation}\label{hww}
2 \ri H^{ABA'B'}  \Omega^{CD}{}_{AA'} \Omega_{CDBB'}  \qquad
  \begin{gathered}
\begin{fmfgraph*}(100,100)
\fmftop{i1,i2}
\fmfbottom{i3}
\fmf{dbl_wiggly}{i1,v1}
\fmf{dbl_wiggly}{i2,v1}
\fmf{dbl_plain}{i3,v1}
\fmflabel{1}{i1}
\fmflabel{2}{i2}
\fmflabel{3}{i3}
\end{fmfgraph*}
\end{gathered} \, ,
\end{equation}

\bigskip\noindent where again the factor of the imaginary unit is the one in front of the action. As we already discussed, all copies of $\Omega$ can be replaced by their expressions (\ref{Omega-H*}). This gives rise to another effective $HHH$ vertex, where now there are two $H$ legs that came from $\Omega$, which needs a decoration by two cilia 
\begin{equation}\label{main-vertex-other}
2 \ri H^{ABA'B'}  \partial_{AR'} H^{\bullet\,CDR'}{}_{A'} \partial_{BS'} H^\bullet_{CD}{}^{S'}{}_{B'} \qquad 
\begin{gathered}
\begin{fmfgraph*}(100,100)
\fmftop{i1,i2}
\fmfbottom{i3}
\fmf{dotted}{v1,i1}
\fmf{dotted}{v1,i2}
\fmf{dbl_plain}{i3,v1}
\fmflabel{1}{i1}
\fmflabel{2}{i2}
\fmflabel{3}{i3}
\end{fmfgraph*}
\end{gathered} \, .
\end{equation}
It is symmetric in $H^\bullet$ placeholders.

\section{Amplitudes}
\label{sec:amplitudes}

\subsection{States}

From the fields $H,\Omega,h,\omega$ only the pair $H,\Omega$ can be non-zero on-shell. The fields $h,\omega$ only describe the scalar part of the metric perturbation, and vanish on-shell. 

For the metric perturbation described by $H$, we take the following usual states 
\begin{equation}
    \epsilon_-^{ABA'B'} =\frac{q^A q^B k^{A'} k^{B'}}{\langle q k\rangle^2}, \qquad \epsilon_+^{ABA'B'} =\frac{k^A k^B q^{A'} q^{B'}}{[qk]^2}.
\end{equation}
Here $k^A, k^{A'}$ are the momentum spinors with the null momentum of the particle being $ k_{AA'}=k_A k_{A'}$. The spinors $q_A, q_{A'}$ are the auxiliary spinors. Here, $\langle q k \rangle = q^A k_A$ and $[q k] = q_{A'} k^{A'}$.

To determine the states for the connection field $\Omega$ we note that on-shell the connection is given by (\ref{Omega-H*}). Substituting here the helicity states for $H$, with the momentum space rule for the derivative being $\partial_{AA'}\to \ri k_{A} k_{A'}$, we see that the connection can only support the positive helicity state
\begin{equation}\label{omega-state}
    \epsilon_+^{ABCA'} = \ri \frac{k^A k^B k^C q^{A'}}{[qk]} \, .
\end{equation}
However, a more efficient version of the Feynman rules consists in replacing $\Omega$ by its expression (\ref{Omega-H*}) everywhere, as we already discussed.

\subsection{Amplitude $--+$}

Three-point amplitudes vanish in the physical Lorentzian signature, but are non-vanishing in, e.g., $(2,2)$ signature. These amplitudes are particularly simple as they are completely fixed by Lorentz invariance, modulo an overall coefficient. It is interesting to reproduce them by our formalism. 

Let us start with an easier $(--+)$ amplitude, where our convention is that the minus helicity is the one that is preferred by our chiral formalism. This is because the connection can only carry the other, positive helicity. Thus, for the amplitude $(--+)$ only the vertex of the type $HH\partial\Omega$ can contribute, with the two negative helicity states being inserted into the $H$ legs. 

With this in mind, we take the usual spinor helicity states for the negative gravitons that we label $1,2$
\begin{equation}
\epsilon^{AA'BB'}_1 = \frac{q^A q^B 1^{A'} 1^{B'}}{\langle q 1\rangle^2},
\quad 
    \epsilon^{AA'BB'}_2 = \frac{q^A q^B 2^{A'} 2^{B'}}{\langle q 2\rangle^2}.
\end{equation}
Here the notation is $k_1^A \equiv 1^A$, and the same for the primed index spinors. If we use the effective version of the Feynman rules with only $H$ field, the positive helicity state to be inserted in the third leg is 
\begin{equation}
    \epsilon_3^{ABA'B'} = \frac{3^A 3^B q^{A'}q^{B'}}{[q3]^2}.
\end{equation}

We now insert the states $1,2$ into the $H$ legs, and state $3$ into $H^\bullet$ leg. The vertex is not explicitly symmetric in the two factors of $H$, and so there will be two contributions. We get
\begin{align}\label{mmp-calc}
2  \ri \left( \frac{q^A q^S 1_{M'} 1^{A'}}{\langle q1\rangle^2}\frac{q^B q^R 2^{M'} 2^{S'}}{\langle q2\rangle^2}+ \frac{q^A q^S 2_{M'} 2^{A'}}{\langle q2\rangle^2}\frac{q^B q^R 1^{M'} 1^{S'}}{\langle q1\rangle^2}\right) 3_R 3_{A'}  3_S 3^{R'} \frac{3_A 3_B q_{R'} q_{S'}}{[q3]^2} \nonumber \\
   =  2\ri \frac{\langle q3\rangle^4 [12]}{\langle q1\rangle^2\langle q2\rangle^2 [q3]} \left( 1^{A'} 2^{S'}-2^{A'} 1^{S'} \right) 3_{A'} q_{S'}= \frac{2}{\ri} \frac{\langle q3\rangle^4 [12]^2}{\langle q1\rangle^2\langle q2\rangle^2} \, .
\end{align}
Using now the momentum conservation in the form
\begin{equation}
    \langle q1\rangle 1^{A'} + \langle q2\rangle 2^{A'} +\langle q3\rangle 3^{A'} =0 \, ,
\end{equation}
we have $\langle q3\rangle/\langle q1\rangle = - [21]/[23]$, $\langle q3\rangle/\langle q2\rangle = - [12]/[13]$, and so we have for this amplitude
\begin{equation}
   \ri  {\cal M}^{--+}= 2\, \frac{[12]^6}{[13]^2[23]^2} \, ,
\end{equation}
which, after restoring the factors of the Newton's constant, becomes the correct answer. 

\subsection{Amplitude $-++$}

Let us now consider the $(-++)$ amplitude. We will take the negative helicity state to correspond to momentum 1, and the two positive helicity states are $2, 3$. Now the vertex $HH\Omega$ can in principle also give contribution, and we will see that it does. 

Let us start with the contribution from (\ref{main-vertex}). The $H^\bullet$ leg can only take one of the positive helicities. However, there are now in total four different possible ways to insert the states. When inserted into an $H$ leg, the positive states correspond to the following helicity spinors:
\begin{equation}
    \epsilon_2^{AA'BB'} = \frac{2_A 2_B q_{A'} q_{B'}}{[q2]^2} \, , \qquad
    \epsilon_3^{AA'BB'} = \frac{3_A 3_B q_{A'} q_{B'}}{[q3]^2}\, .
\end{equation}
Let us start with the two terms that get generated by inserting $1,2$ into $HH$ and $3$ into $H^\bullet$. This is similar to (\ref{mmp-calc}): 
\begin{align}\label{mmp-calc-1}
 2\ri  \left( \frac{q^A q^S 1_{M'} 1^{A'}}{\langle q1\rangle^2}\frac{2^B 2^R q^{M'} q^{S'}}{[q2]^2}+ \frac{2^A 2^S q_{M'} q^{A'}}{[q2]^2}\frac{q^B q^R 1^{M'} 1^{S'}}{\langle q1\rangle^2}\right) 3_R 3_{A'} 3_S 3^{R'} \frac{3_A 3_B q_{R'} q_{S'}}{[q3]^2} \nonumber \\
  =  \frac{2}{\ri} \frac{[q1]^2 \langle 23\rangle^2 \langle q3\rangle^2}{\langle q1\rangle^2[q2]^2} \, ,
\end{align}
where now only the second term in the brackets contributes. We now add to this the $2,3$ permutation
\begin{equation}\label{mpp-1}
   \frac{2}{\ri} \frac{[q1]^2 \langle 23\rangle^2 }{\langle q1\rangle^2} \left( \frac{\langle q3\rangle^2}{[q2]^2}+ \frac{\langle q2\rangle^2}{[q3]^2}\right) \, .
\end{equation}

We now evaluate a contribution from the vertex (\ref{main-vertex-other}). In order to be able to recycle the result in later computations of the 4 point amplitudes, let us evaluate this vertex by inserting the states $2,3$ into the $H^\bullet$ legs, and a placeholder field $H_{ABA'B'}$ into the $H$ leg. We have
\begin{equation}
	-2\ri H^{ABA'B'} 2_A 2_{R'} \frac{2^C 2^D q^{R'} q_{A'}}{[q2]^2} 3_B 3_{S'} \frac{3_C 3_D q^{S'} q_{B'}}{[q3]^2} = \frac{2}{\ri} H_{AB}{}^{R'S'} 2^A 3^B q_{R'} q_{S'} \frac{ \langle 23\rangle^2}{ [q2][q3]} \, ,
\end{equation}
where the minus sign in front of the first expression comes from the two derivatives in the vertex. We need to add to this the same expression with $2,3$ interchanged, which gives 
\begin{equation}\label{mpp-placeholder}
\frac{2}{\ri} 
    H_{AB}{}^{R'S'} \left( 2^A 3^B + 3^A 2^B \right) q_{R'} q_{S'} 
    \frac{ \langle 23\rangle^2}{[q2][q3]}\, .
\end{equation}

We now evaluate this inserting the negative helicity state of the graviton 1 instead of the placeholder. We get
\begin{equation}\label{mpp-2}
\frac{4}{\ri} \frac{\langle q2\rangle \langle q3\rangle [q1]^2 \langle 23\rangle^2}{\langle q1\rangle^2 [q2][q3]}\, .	
\end{equation}
Adding (\ref{mpp-1}) and twice (\ref{mpp-2}), we get
\begin{align}\label{mpp-3}
   \frac{2}{\ri} \frac{[q1]^2 \langle 23\rangle^2 }{\langle q1\rangle^2[q2]^2[q3]^2} \left( \langle q2\rangle^2 [q2]^2 + \langle q3\rangle^2 [q3]^2 + 2 \langle q2\rangle \langle q3 \rangle [q2] [q3]\right) \nonumber \\
    =\frac{2}{\ri} \frac{[q1]^2 \langle 23\rangle^2 }{\langle q1\rangle^2[q2]^2[q3]^2} \left( \langle q2\rangle [q2] + \langle q3\rangle [q3] \right)^2 = \frac{2}{\ri}  \frac{[q1]^4 \langle 23\rangle^2 }{[q2]^2[q3]^2}\, , 
\end{align}
where we have used the momentum conservation in the form
\begin{equation}
    \langle q1\rangle [q1]+ \langle q2\rangle [q2] + \langle q3\rangle [q3]=0 \, .
\end{equation}
We can now use the momentum conservation to convert the square brackets in (\ref{mpp-3}). Altogether, this gives the correct answer for the amplitude:
\begin{equation}
    \ri {\cal M}^{-++} = 2\, \frac{\langle 23\rangle^6}{\langle 12\rangle^2 \langle 13\rangle^2}\, .
\end{equation}
This computation shows that the vertex (\ref{hww}) is essential for getting the right answer for the amplitudes. 

\section{Currents}
\label{sec:currents}

The technology of currents is very efficient as it works by computing objects that are later recycled in other calculations. A current is an object obtained as the sum of all Feynman diagrams with every except one leg on shell. 

\subsection{$J(1^-, 2^-)$ current}

Let us start by computing the simplest negative-negative current obtained by inserting two negative gravitons states into the $H$ legs of the vertex (\ref{main-vertex-prelim}), followed by the $\Omega H$ propagator. Equivalently, we use the effective vertex (\ref{main-vertex}). This gives the current that we denote as $J^{ABA'B'}(1^-,2^-)$. The insertion of two negative helicity gravitons $1,2$ into the vertex (\ref{main-vertex}), followed by the $HH$ propagator, is given by
\begin{align*}
 J^{ABR'S'}(1^-,2^-)=\frac{2}{(1+2)^2}  \left( \frac{q^A q^S 1_{M'} 1^{A'}}{\langle q1\rangle^2}\frac{q^B q^R 2^{M'} 2^{S'}}{\langle q2\rangle^2}+ \frac{q^A q^S 2_{M'} 2^{A'}}{\langle q2\rangle^2}\frac{q^B q^R 1^{M'} 1^{S'}}{\langle q1\rangle^2}\right) \\ 
 {} \times (1+2)_{RA'} (1+2)_{S}{}^{R'}
 = - \frac{2}{(1+2)^2} \frac{q^A q^S q^B q^R [12]^2}{\langle q1\rangle^2 \langle q2\rangle^2}  (1+2)_{R}{}^{S'} (1+2)_{S}{}^{R'} \, .
 \end{align*}
We now replace $(1+2)^2=2\langle 12\rangle [12]$ and get 
\begin{equation}\label{current}
J^{ABA'B'}(1^-,2^-)=q^A  q^B \langle q| 1+2|^{A'} \langle q| 1+2|^{B'} J(1^-,2^-)\, ,
\end{equation}
where we have introduced the notation
\be
 \langle q| 1+2|^{R'} = q^{R} (1+2)_R{}^{R'},
 \ee
as well as the current with its index structure stripped off:
\begin{equation}\label{mm-current}
J(1^-,2^-):=- \frac{  [12]}{\langle q1\rangle^2 \langle q2\rangle^2 \langle 12\rangle}\, .
\end{equation}

We note that the single negative helicity state can also be written in the form (\ref{current}). Indeed, we can write
\begin{equation}
\epsilon_-^{ABA'B'} \equiv J^{ABA'B'}(1^-) = q^A  q^B \langle q| 1|^{A'} \langle q| 1|^{B'} J(1^-)\, ,
\end{equation}
where
\begin{equation}
J(1^-)=\frac{1}{\langle q1\rangle^4} \, .
\end{equation}

The currents $J(1^-)$, $J(1^-,2^-)$ are the beginning of a sequence of all negative helicity currents, for which a recursion relation can be written and solved in a closed form; see, e.g., \cite{Krasnov:2013wsa} for the recursion as well as the general expression for this current.  

\subsection{$J(1^-,2^+)$ current}

We now repeat the previous current calculation, but this time consider the coupling of a negative and a positive helicity states. We first consider contribution of the vertex (\ref{main-vertex}), and first compute the insertion of $1^-,2^+$ into the $H$ legs of this vertex. We get
\begin{align*}
 J^{ABR'S'}(1^-,2^+)=\frac{2}{(1+2)^2}  \left( \frac{q^{(A} q^S 1_{M'} 1^{A'}}{\langle q1\rangle^2}\frac{2^{B)} 2^R q^{M'} q^{S'}}{[q2]^2}+ \frac{2^{(A} 2^S q_{M'} q^{A'}}{[q2]^2}\frac{q^{B)} q^R 1^{M'} 1^{S'}}{\langle q1\rangle^2}\right) \\ 
{} \times (1+2)_{RA'} (1+2)_{S}{}^{R'}
 =  \frac{2^{(A} q^{B)} \langle q| 1+2 | q] [q1]}{\langle q1\rangle^2 [q2]^2 [12]}  1^{R'} 1^{S'}.
 \end{align*}
 We note that this expression can be put into the form (\ref{current}) if we choose the gauge $q^A=2^A$, in which the auxiliary spinor of the negative helicity gravitons is equal to the momentum spinor of the single positive helicity graviton. Indeed, in this case, the current becomes
 \begin{equation}
 J^{ABA'B'}(1^-,2^+) = q^A  q^B \langle q| 1+2|^{A'} \langle q| 1+2|^{B'} J(1^-,2^+)\, ,
\end{equation}
where
\begin{equation}\label{mp-current}
J(1^-,2^+) = - \frac{[q1]^2}{\langle q1\rangle^2 [q2]^2 [12]\langle 12\rangle}\, .
\end{equation}
Note that this only exhibits the correct scaling property with respect to graviton $1$. The degree of homogeneity for the graviton $2$ is now partially carried by the factors of $q$ in the spinor prefactor. 

Let us now consider the case where the $2^+$ graviton is inserted into the $H^\bullet$ leg of (\ref{main-vertex}) instead. Inserting the $1^-$ into one of the two $H$ legs, and keeping the other $H$ as a placeholder, we get
\begin{equation}\label{mp-diff-insert}
- \frac{2}{(1+2)^2} \left(\frac{ q^A q^S 1_{M'} 1^{R'}}{\langle q1\rangle^2} H^{BR}{}^{M'S'} +H^{AS}{}_{M'}{}^{R'} \frac{q^B q^R 1^{M'} 1^{S'}}{\langle q1\rangle^2} \right) 2_R 2_{R'} 2_{S} 2_{K'} \frac{2_A 2_B q^{K'} q_{S'}}{[q2]^2} \, .
\end{equation}
By using the Schouten identity, this gives the following contribution to the current:
\begin{equation}
- \frac{2 \langle q2\rangle^2}{(1+2)^2 \langle q1\rangle^2} 2^A 2^B 1^{A'} 1^{B'}\, .
\end{equation}
This vanishes in our gauge $q^A=2^A$. 

There is another contribution to this current, obtained by inserting a positive and a negative helicity state into (\ref{main-vertex-other}). We must insert the negative state into the $H$ leg of this vertex, while the positive state can go into either of the two $H^\bullet$ legs. This gives 
\begin{equation}
\frac{4}{(1+2)^2} \frac{ \tbr{q2} \sbr{q1}}{\tbr{q1}^2 \sbr{q2}} 2^A 2^B \left( \tbr{q1} 1^{A'} + \tbr{q2} 2^{A'} \right)  1^{B'} \, ,
\end{equation}
which again vanishes in the gauge $q^A=2^A$, and so this contribution can be ignored in this gauge.

 \subsection{$J(1^+,2^+)$ current}
 
Although for the computations that follow we do not need this current, let us also consider the current of two positive helicity states, for completeness. Let us start with the contribution from (\ref{main-vertex}). The insertion of both states into the $H$ legs vanishes because the auxiliary spinors $q^{A'}$ get contracted. Thus, we must insert one of the states into the $H^\bullet$ leg. Taking this to be the state $2$, we get a modification of (\ref{mp-diff-insert}):
\begin{equation}
- \frac{2}{(1+2)^2} \left(\frac{ 1^A 1^S q_{M'} q^{R'}}{[q1]^2} H^{BR}{}^{M'S'} +H^{AS}{}_{M'}{}^{R'} \frac{1^B 1^R q^{M'} q^{S'}}{[q1]^2} \right) 2_R 2_{R'} 2_{S} 2_{K'} \frac{2_A 2_B q^{K'} q_{S'}}{[q2]^2}\, .
\end{equation}
Only the first term contributes, and we get the following contribution to this current:
\begin{equation}
- \frac{2 \langle 12\rangle^2}{(1+2)^2 [q1]^2} 2^A 2^B q^{A'} q^{B'}\, .
\end{equation}
This must be symmetrised with respect to $1\leftrightarrow 2$. 

Let us now consider the contribution of the vertex (\ref{main-vertex-other}). The only non-vanishing option is to insert both positive states into the $H^\bullet$ legs. We then have the following contribution: 
\begin{equation}
- \frac{2\langle 12\rangle^2}{(1+2)^2 [q1][q2]} (1^A 2^B + 2^A 1^B) q^{A'} q^{S'}\, .
\end{equation}
Collecting all contributions, we get
\begin{equation}
J(1^+,2^+)^{ABA'B'} = - \frac{\langle 12\rangle}{[12] [q1]^2[q2]^2} [q|1+2|^A [q|1+2|^B q^{A'} q^{B'}\, ,
\end{equation}
which is just the complex conjugate of (\ref{current}).

\subsection{Coupling current to a negative helicity state}

We now compute the result of coupling a current of the general form (\ref{current}) to a negative helicity state, inserting both into the vertex (\ref{main-vertex}) and then following by the propagator. We have
\begin{align}
J^{ABR'S'}(K,3^-)=\frac{2}{( K+3)^2} \left( J^{(A|S|}{}_{M'}{}^{A'}(K) \frac{q^{B)} q^R 3^{M'} 3^{S'}}{\langle q3\rangle^2}+ \frac{q^{(A} q^{|S|} 3_{M'} 3^{A'}}{\langle q3\rangle^2} J^{B)RM'S'}(K)\right) \nonumber \\ {} \times (K+3)_{RA'} (K+3)_S{}^{R'} \nonumber \\ =
\frac{2 J(K)}{( K+3)^2} \left( q^A q^S \langle q|K|_{M'} \langle q| K|^{A'}\frac{q^B q^R 3^{M'} 3^{S'}}{\langle q3\rangle^2}+ \frac{q^A q^S 3_{M'} 3^{A'}}{\langle q3\rangle^2} q^B q^R \langle q| K|^{M'} \langle q|K|^{S'}\right) \nonumber \\ {} \times (K+3)_{RA'} (K+3)_S{}^{R'}\, ,
\end{align}
where $K$ is the sum of all the momenta of particles participating in the current $J^{ABA'B'}(K)$, and $J (K)$ is the scalar part of this current, see (\ref{current}). A computation similar to one performed for $J(1^-,2^-)$ shows that the current obtained by coupling a smaller current to a negative helicity graviton keeps its form
\begin{equation}
J^{ABR'S'}(K,3^-) = q^A q^B \langle q| K+3|^{R'} \langle q| K+3|^{S'} J(K,3^-)\, ,
\end{equation}
where
\begin{equation}
J(K,3^-) = - \frac{2}{(K+3)^2 \langle q3\rangle^2} J(K) \langle q| K|3]^2\, .
\end{equation}
For $K=1^-$, and $3$ replaced by $2$, this reproduces the previous result (\ref{mm-current}).

\subsection{Coupling current to a positive helicity state}

We now do a similar computation but now couple a current of the general form (\ref{current}) to a positive helicity state. We have
\begin{align}
J^{ABR'S'}(K,3^+)=\frac{2}{( K+3)^2} \left( J^{(A|S|}{}_{M'}{}^{A'}(K) \frac{3^{B)} 3^R q^{M'} q^{S'}}{[q3]^2}+ \frac{3^{(A} 3^{|S|} q_{M'} q^{A'}}{[q3]^2} J^{B)RM'S'}(K)\right) \nonumber \\ {} \times (K+3)_{RA'} (K+3)_S{}^{R'} \nonumber \\ =
\frac{2 J(K)}{( K+3)^2} \left( q^{(A} q^{|S|} \langle q|K|_{M'} \langle q| K|^{A'}\frac{3^{B)} 3^R q^{M'} q^{S'}}{[q3]^2}+ \frac{3^{(A} 3^{|S|} q_{M'} q^{A'}}{[q3]^2} q^{B)} q^R \langle q| K|^{M'} \langle q|K|^{S'}\right) \nonumber \\ {} \times (K+3)_{RA'} (K+3)_S{}^{R'}.
\end{align}
If we now assume that the current inserted is an all-negative current, and assume that the negative helicity auxiliary spinors $q^A$ are equal to the momentum spinor of the positive helicity graviton $q^A=3^A$, then the first term does not contribute. This is because of the combination
\begin{equation}
\langle q| K|^{A'} \langle 3|K+3|_{A'} = \langle q| K|^{A'} \langle 3|K|_{A'} \sim K^2 \langle q3\rangle\, , 
\end{equation}
which vanishes if $q=3$. The second term gives, in the same gauge,
\begin{equation}
J^{ABR'S'}(K,3^+) = q^A q^B \langle q| K+3|^{R'} \langle q| K+3|^{S'} J(K,3^+)\, ,
\end{equation}
where
\begin{equation}\label{Kp-current}
J(K,3^+) = - \frac{2J(K) \langle q| K+3|q]^2}{(K+3)^2 [q3]^2}\, .
\end{equation}
Again, for $J(K)$ being $J(1^-)$ and $3=2$, this reproduces the previous result (\ref{mp-current}).

The contribution to the coupling of a current to a positive helicity graviton $3$ via the vertex (\ref{main-vertex-other}) is also checked to vanish in the gauge $q^A=3^A$, so (\ref{Kp-current}) is the complete answer in this gauge. 

\subsection{Two negative one positive helicity current}

Let us use the above results to evaluate the $J(1^-,2^-,3^+)$ current. There are three terms that are contributing. One can couple $J(1^-,2^-)$ to $3^+$, as well as $J(1^-,3^+)$ to $2^-$ and $J(2^-,3^+)$ to $1^-$. This gives
\begin{equation}\label{mmp-current-index}
J^{ABA'B'}(1^-,2^-,3^+) = q^A q^B \langle q| 1+2+3|^{A'} \langle q| 1+2+3|^{B'} J(1^-,2^-,3^+) \, ,
\end{equation}
where
\begin{equation}
J(1^-,2^-,3^+) = \frac{2  [12]^2}{(1+2+3)^2 \langle q1\rangle^2 \langle q2\rangle^2 [q3]^2} \left( \frac{\langle q| 1+2|q]^2}{\langle 12\rangle [12]}  +
 \frac{[q1]^2\langle q1\rangle^2}{[13]\langle 13\rangle} +  \frac{[q2]^2\langle q2\rangle^2}{[23]\langle 23\rangle}  \right) \, .
\end{equation}
The expression in the brackets here can be written as
\begin{align*}
[q1]^2\langle q1\rangle^2 \left( \frac{1}{\langle 12\rangle [12]} +  \frac{1}{\langle 13\rangle [13]} \right) + 
[q2]^2\langle q2\rangle^2 \left( \frac{1}{\langle 12\rangle [12]} +  \frac{1}{\langle 23\rangle [23]} \right) +
\frac{2 \langle q1\rangle \langle q2\rangle [q1] [q2] }{\langle 12\rangle [12]} \\ 
= \frac{[q1]^2\langle q1\rangle^2 ( K^2/2- \langle 23\rangle [23])}{\langle 12\rangle [12]\langle 13\rangle [13]} + \frac{[q2]^2\langle q2\rangle^2 ( K^2/2- \langle 13\rangle [13])}{\langle 12\rangle [12]\langle 23\rangle [23]} + \frac{2 \langle q1\rangle \langle q2\rangle [q1] [q2] }{\langle 12\rangle [12]} \, .
\end{align*}
Here, 
\begin{equation}
\frac{1}{2} K^2 = \frac{1}{2} (1+2+3)^2 = \langle 12\rangle [12]+\langle 13\rangle [13]+\langle 23\rangle [23] \, .
\end{equation}
The above expression has a part proportional to $K^2$, and the part
\begin{align}
&\frac{1}{\langle 12\rangle [12]\langle 13\rangle [13] \langle 23\rangle [23]} \left( - [q1]^2\langle q1\rangle^2 \langle 23\rangle^2 [23]^2 - [q2]^2\langle q2\rangle^2 \langle 13\rangle^2 [13]^2 \right. \nonumber \\ &{} + 2 \left. \vphantom{[q1]^2} \langle q1\rangle \langle q2\rangle [q1] [q2] \langle 13\rangle [13] \langle 23\rangle [23] \right) = 
- \frac{\left( [q1]\langle q1\rangle \langle 23\rangle [23]-[q2]\langle q2\rangle \langle 13\rangle [13] \right)^2}{\langle 12\rangle [12]\langle 13\rangle [13] \langle 23\rangle [23]} \, .
\end{align}
We now use the fact that $q^A=3^A$ in our gauge, as well as the Schouten identity
\begin{equation}
-[q1][23]+[q2][13]=[q3][12] \, .
\end{equation}
This shows that the part not proportional to $K^2$ is given by
\begin{equation}
- \frac{ \langle 13\rangle \langle 23\rangle [12] [q3]^2}{\langle 12\rangle [13][23]}\, .
\end{equation}
Thus, overall, we get
\begin{equation}\label{mmp-current}
J(1^-,2^-,3^+) = \frac{[12]( [q1]^2 \langle 13\rangle [23]+ [q2]^2 \langle 23\rangle [13])}{\langle 12\rangle \langle 13\rangle^2 \langle 23\rangle^2 [13][23] [q3]^2}  - \frac{2 [12]^3}{K^2 \langle 12\rangle \langle 13\rangle \langle 23\rangle [13][23]} \, .
\end{equation}
Note that the term containing $K^2$ in the denominator is gauge invariant, i.e.,  independent of the auxiliary spinor $q^{A'}$. We will use this result to compute the graviton-graviton scattering amplitude. 

It can be shown that the pattern visible for the $1^-,2^-,3^+$ current continues, and a general all except one negative helicity current is of the form (\ref{mmp-current-index}), with its scalar part given by a term where $K^2$ has cancelled, as well as a gauge-invariant (i.e., $q^{A'}$ independent) term where $K^2$ remains in the denominator. One can obtain a recursion relation for the gauge-independent part, see \cite{Delfino:2014xea}, but no closed expression for a solution of this recursion is known. 

\subsection{Contribution of the 4-valent vertex}

The computation above was carried out by taking into account contributions from the cubic vertices. However, already for the current involving 3 gravitons, we may need to consider the contribution from the 4-valent vertex. It is easy to see that this contribution vanishes in the gauge that we used. To see this, we take the $HH\Omega\Omega$ vertex in the form [see (\ref{L3})]
\begin{equation}
	H^{ABA'B'} H^{CD}{}_{A'}{}^{C'} \left( \partial_{DM'} H^\bullet_A{}^{FM'}{}_{B'}  \partial_{CN'} H^\bullet_{BF}{}^{N'}{}_{C'} - \partial^F{}_{M'} H^\bullet_{AB}{}^{M'}{}_{C'} \partial_{DN'} H^\bullet_{CF}{}^{N'}{}_{B'} \right) \,.
\end{equation}
 We then insert two currents of the general form (\ref{mmp-current-index}) into the two $H$ legs, and a positive helicity state into one of the $H^\bullet$ legs. It is easy to see that there are factors of the auxiliary spinor $q^A$ contracting with itself, in the gauge where the $q^A$ of the negative helicity states is equal to the momentum spinor of the positive helicity state. So, the 4-valent vertex does not contribute to computations of the all except one negative currents, in the gauge used. 

\section{Graviton-graviton scattering}
\label{sec:scattering}

\subsection{Vanishing amplitudes}

We now compute the graviton-graviton, or a 4-point amplitude. Let us first convince ourselves that only the $--++$ such amplitude can be non-zero. 

Let us consider the $----$ amplitude first. Such states can only be inserted into the field $H$ external legs. However, it is easy to see that one cannot construct a 4-point tree level diagram with four external legs being those of the $H$ field. So, this amplitude must vanish.

Let us consider the $---+$ amplitude. We now can construct a diagram with 3 external $H$ legs and one external $\Omega$, for example
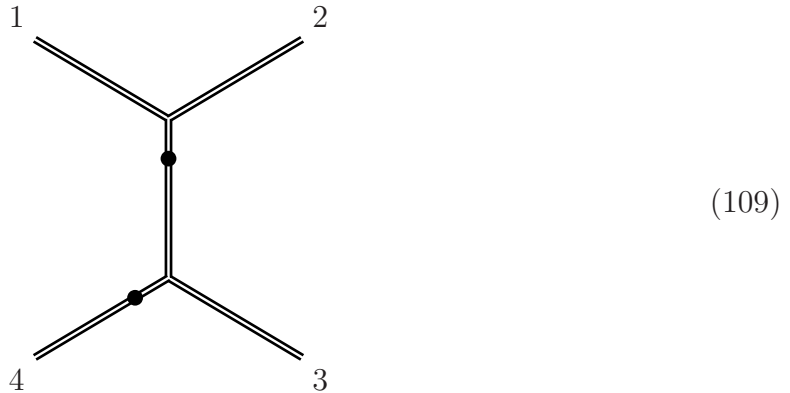
\begin{equation}\label{diag-mmp}
\begin{gathered}	
\begin{fmfgraph*}(100,150)
\fmftop{i1,i2}
\fmfbottom{i4,i3}
\fmf{dbl_plain}{i1,v1}
\fmf{dbl_plain}{i2,v1}
\fmf{dbl_plain}{i3,v2}
\fmf{dotted}{v2,i4}
\fmf{dotted}{v1,v2}
\fmflabel{1}{i1}
\fmflabel{4}{i4}
\fmflabel{2}{i2}
\fmflabel{3}{i3}
\end{fmfgraph*}
\end{gathered}
\end{equation}
Thus, we must analyse the situation more closely. We will use the effective version of the vertices, where there is always two derivatives in the vertex, and no derivatives in the propagator. The 3 helicity states for $H$ carry six copies of the auxiliary spinor $q^A$. We choose this spinor to be equal to the momentum spinor $k^A$ of the fourth positive helicity state. In total, there are 8 copies of the spinor $q^A$. These must somehow contract by the spinor metrics present in the vertices and the propagators. Any contraction of a pair of such spinors gives zero, so they cannot contract between themselves, and can only contract into the unprimed spinor indices of the factors of momenta that come from the derivatives in the vertices. There are 4 such derivatives in this diagram. There are simply not enough factors of momenta to give a non-zero result. 

Exactly the same logic applies to the amplitudes $-+++$ and $++++$, but this time applied to primed auxiliary spinors. 

\subsection{The non-vanishing $--++$ amplitude}

Let us now compute the non-vanishing $--++$ amplitude. We label the gravitons so that the gravitons $1,2$ are negative helicity, and $3,4$ are positive helicity. 

We will compute this amplitude by utilising the previous results on the currents, specifically the result (\ref{mmp-current-index}) and (\ref{mmp-current}) on the current of 2 negative helicity gravitons $1,2$ and a third positive helicity graviton $3$. We will then put this current on-shell, which extracts the 4-point amplitude. 

Given the result (\ref{mmp-current}), the 4-point amplitude is easily computed. First, we need to amputate the propagator from the off-shell leg of the current, i.e., multiply the result by $\ri K^2$. Given that $K^2=4^2=0$, this kills the first term in (\ref{mmp-current}). Thus, the second term in (\ref{mmp-current}) is essentially the result. We just need to correct it by the prefactor that comes by inserting the positive helicity state $4^A 4^B q^{A'} q^{B'}/[q4]^2$ into the current (\ref{mmp-current-index}). The factors of the auxiliary spinor $q^{A'}$ cancel out, and the amplitude is given by the last term in (\ref{mmp-current}) multiplied by $\ri K^2$, and multiplied by $\langle 34\rangle^4$. This gives
\begin{equation}
{\cal M}^{--++} = 2\ri \frac{\langle 34\rangle^6}{ \langle 13\rangle \langle 23\rangle \langle 14\rangle \langle 24\rangle } \frac{[12]}{\langle 12\rangle}\, , 
\end{equation}
where we have used the momentum conservation in the form $[12]/[13]=-\langle 34\rangle/\langle 24\rangle$ and $[12]/[23] = \langle 34\rangle /\langle 14\rangle$ to convert square brackets in the denominator into the angle brackets. This is the correct result for the graviton-graviton amplitude. 

\section{Conclusions}
\label{sec:conclusions}

In this paper, starting from the chiral action (\ref{action}) of first-order formalism for GR in two-component spinor notation and then fixing the diffeomorphism and Lorentz-group gauge freedom, we finally arrived at a remarkably simple free part (\ref{Lnew}) of the Lagrangian. It describes two sectors of fields $(\Omega^{ABCA'}, H^{ABA'B'})$ and $(\omega^{AA'}, h^{AB})$ that decouple from each other and, moreover, have absent $\Omega$--$\Omega$ and $\omega$--$\omega$ propagators [see the structure of propagators in (\ref{propag})].  The pair $(\Omega^{ABCA'}, H^{ABA'B'})$ can describe gravitons in asymptotic states, while the pair $(\omega^{AA'}, h^{AB})$ can only be present in the propagators of internal lines.  The fields $\Omega^{ABCA'}$ and $\omega^{AA'}$ are composed of the spin connection one-form $\omega^{AB}$ and the Lagrange multipliers that fix the diffeomorphism gauge freedom. The fields $H^{ABA'B'}$ and $h^{AB}$ are composed of the perturbation $h^{AA'}$ of the tetrad one-form and the Lagrange multipliers that fix the gauge freedom of the chiral half of the Lorentz group.

Fixing gauge freedom can be extended to include also the interaction part of the action.  However, this can be done in a number of non-equivalent ways, and the simplest and most useful form still remains to be found.  Without making this choice, it is already possible to apply the formalism to various computational problems, and we illustrated it by calculating the simplest amplitudes, such as 3- and 4-point graviton scattering amplitudes. For these simplest problems only the fields $(\Omega^{ABCA'}, H^{ABA'B'})$ matter, and moreover, only the vertices (\ref{main-vertex}) and (\ref{main-vertex-other}) contribute. The reader will hopefully appreciate the ease with which the formalism we developed reproduces the known results. 

The Feynman rules that result from a first-order formalism such as one we consider in this paper deal with both metric and connection fields. However, in the absence of the connection-connection propagator, there exists an effective version of the Feynman rules in which the derivative present in the numerator of the tetrad-connection propagator is assigned to the vertex. This is done by replacing the $\Omega$ in vertices by derivatives of $H$, see (\ref{Omega-H*}). In this effective version of the perturbation theory there is only the tetrad field that propagates, with the propagator given by (\ref{prop-HH}), and all vertices contain two copies of the derivative operator. However, one has to mark the vertex legs that came from the connection by appropriate cilia, and contraction of two legs with cilia are forbidden in view of the absence of the connection-connection propagator. The formalism that arises this way is quite similar to that in the case of the chiral Yang--Mills theory.

There are other interesting calculations that can be performed with the formalism we developed. First, it is not hard to generalise our perturbative expansion and gauge-fixing to backgrounds other than Minkowski. The only change in this case is that the partial derivative operator needs to be replaced by the appropriate covariant derivative operator with respect to the background connection. It would be interesting to apply this to a computation of the one-loop effective action on a general Einstein background. Also, given the simplicity of the propagators and vertices in this formalism, it is possible that even a chiral version of the two-loop computation \cite{Goroff:1985th} may be within reach. 

\section*{Acknowledgements}

The work of Y.~S. was supported by the National Academy of Sciences of Ukraine (project 0116U003191) and by the scientific program ``Astronomy and Space Physics'' (project 19BF023-01) of the Taras Shevchenko National University of Kiev.

\section{Appendix: Alternative derivation}

In the main text, we have motivated our gauge-fixing procedure and the introduction of fields $(\Omega^{ABCA'}, H^{ABA'B'})$ and $(\omega^{AA'}, h^{AB})$ using  different spinor representations (\ref{orP}),  (\ref{orW}) of the totally anti-symmetric $\epsilon$ tensor. We now sketch an alternative path, where most of the manipulations needed to arrive at the final result are done already at the level of the expansion of the $\Sigma^{AB}$ 2-form. 

\subsection*{Expansion of the SD 2-form}

The $\Sigma^{AB}$ 2-form in the action has the following expansion:
\begin{align}\label{sigma-pert}
\Sigma^{AB} &= \frac12 \left( \theta^A{}_{C'}+ h^A{}_{C' MM'} \theta^{MM'} \right) \wedge \left( \theta^{BC'} +h^{BC'}{}_{NN'} \theta^{NN'} \right) \nonumber \\ &=
\frac12 \left( \vphantom{\epsilon^{A'}{}_{M'}} \epsilon^A{}_M \epsilon_{C'M'} + h^A{}_{C' MM'} \right) \left( \epsilon^B{}_N \epsilon^{C'}{}_{N'} +h^{BC'}{}_{NN'} \right) \theta^{MM'}\wedge \theta^{NN'} \nonumber \\ &=
\frac12 \left[ \epsilon^A{}_M \epsilon^B{}_N \epsilon_{M'N'} + 2 \epsilon^{(A}{}_M h^{B)}{}_{M'NN'} + h^A{}_{C'MM'} h^{BC'}{}_{NN'} \right] \theta^{MM'}\wedge \theta^{NN'} \, .
\end{align}

We now compute this in the parametrisation (\ref{hdeco}). We have
\begin{align}
\Sigma^{AB} &= \frac{1}{2} \left[\vphantom{\epsilon^{A'}{}_{M'}} \epsilon^A{}_M \epsilon^B{}_N \epsilon_{M'N'} + 2 \epsilon^{(A}{}_M h^{B)}{}_{NM'N'} + 2 \epsilon^{(A}{}_M h^{B)}{}_N \epsilon_{M'N'} \right. \nonumber \\ &\left. {} + h^A{}_{MC'M'} h^B{}_N{}^{C'}{}_{N'} + 2 h^{(A}{}_M h^{B)}{}_{NM'N'}  + h^A{}_M h^B{}_N \epsilon_{M'N'} \right] \theta^{MM'}\wedge \theta^{NN'} \, .
\end{align}
Collecting the terms with $\epsilon_{M'N'}$ we get
\begin{align} \label{sigma-pert*}
\Sigma^{AB} &= \left( \frac{1}{2} \left[ \epsilon^{(A}{}_M + h^{(A}{}_M \right] \left[\epsilon^{B)}{}_N +  h^{B)}{}_N \right] \epsilon_{M'N'} + \left[ \epsilon^{(A}{}_M + h^{(A}{}_M \right] h^{B)}{}_{NM'N'} \right. \nonumber \\ &\left. {} + \frac{1}{2} h^A{}_{MC'M'} h^B{}_N{}^{C'}{}_{N'} \right) \theta^{MM'}\wedge \theta^{NN'}.
\end{align}
The first term here is SD, the second is ASD, and the third contains both parts. It is interesting that the object $h_{AB}$ always appears in the combination with $\epsilon_{AB}$.

\subsection*{Computation}

We now establish a convenient form of the 2-form perturbation wedged with two copies of the background tetrad. Thus, we now wedge (\ref{sigma-pert*}) with $\theta^{RR'}\wedge\theta^{SS'}$ and use (\ref{orW}) to get 
\begin{align}\label{comp-1}
\frac{1}{\ri} \Sigma^{AB}\wedge \theta^{RR'}\wedge \theta^{SS'} &= \left[ \epsilon^{(A|R|} + h^{(A|R|} \right] \left[\epsilon^{B)S} + h^{B)S} \right]  \epsilon^{R'S'} + \left[ \epsilon^{(A|M|} + h^{(A|M|} \right] h^{B)}{}_{M}{}^{R'S'} \epsilon^{RS}  \nonumber \\ &\quad {} + \frac{1}{2} h^{(A|R|}{}_{M'N'} h^{B)S}{}^{M'N'} \epsilon^{R'S'} - \frac{1}{2} h^{(A}{}_{MM'}{}^{|R'|} h^{B)MM'S'} \epsilon^{RS} \, ,
\end{align}
where we have omitted the volume form $\upsilon$ on the right-hand side. Expanding the brackets in the first term here we have
\begin{equation}
\epsilon^{(A|R|} \epsilon^{B)S} + h^{(A|R|} h^{B)S} + h^{(A|R|} \epsilon^{B)S} + h^{(A|S|} \epsilon^{B)R} \, .
\end{equation}
Using the Schouten identity, we can rewrite the last two terms as
\begin{equation}
h^{(A|R|} \epsilon^{B)S} + h^{(A|S|} \epsilon^{B)R} = 2 h^{(A|R|} \epsilon^{B)S} - \epsilon^{RS} h^{(AB)} \, .
\end{equation}
The first line in (\ref{comp-1}) then becomes
\begin{align}
&\left[ \epsilon^{(A|R|} \epsilon^{B)S} + h^{(A|R|} h^{B)S} + 2h^{(A|R|} \epsilon^{B)S} \right] \epsilon^{R'S'} \nonumber \\ &{} + \left[ \epsilon^{(A|N|} + h^{(A|N|} \right] h^{B)}{}_{N}{}^{R'S'} \epsilon^{RS} - h^{(AB)} \epsilon^{RS}\epsilon^{R'S'}\, .
\end{align}
We can rewrite the second line as
\begin{equation}
\left[ h^{AB}{}^{R'S'} - h^{(AB)} \epsilon^{R'S'} + h^{(A|N|} h^{B)}{}_{N}{}^{R'S'} \right] \epsilon^{RS} \, .
\end{equation}
Thus, overall, we can rewrite (\ref{comp-1}) as
\begin{align}\label{comp-2}
&\left[ \epsilon^{(A|R|} \epsilon^{B)S} + 2h^{(A|R|} \epsilon^{B)S} \right] \epsilon^{R'S'} +  \left[ h^{AB}{}^{R'S'} - h^{(AB)} \epsilon^{R'S'} \right] \epsilon^{RS} \nonumber \\
&{} + \frac{1}{2} h^{(A|R|}{}_{M'N'} h^{B)S}{}^{M'N'} \epsilon^{R'S'} + h^{(A|R|} h^{B)S} \epsilon^{R'S'} \nonumber \\
&{} - \frac{1}{2} h^{(A}{}_{MM'}{}^{|R'|} h^{B)MM'S'}  \epsilon^{RS}+ h^{(A|N|} h^{B)}{}_{N}{}^{R'S'} \epsilon^{RS} \, .
\end{align}

We note that the first term in the second and third lines can be combined using the Schouten identity
\begin{equation}
\frac{1}{2} h^{(A|R|}{}_{M'N'} h^{B)S}{}^{M'N'} \epsilon^{R'S'}- \frac{1}{2} h^{(A}{}_{MM'}{}^{|R'|} h^{B)MM'S'} \epsilon^{RS} = h^{(A |S}{}_{M'}{}^{R'|} h^{B)R M'S'} \, .
\end{equation}
Thus, an even more compact form of (\ref{comp-1}) is
\begin{align}\label{sigma-expanded}
\frac{1}{\ri} \Sigma^{AB}\wedge \theta^{RR'}\wedge \theta^{SS'} = \left[ \epsilon^{(A|R|} \epsilon^{B)S} + 2h^{(A|R|} \epsilon^{B)S} \right] \epsilon^{R'S'} +  \left[ h^{AB}{}^{R'S'} - h^{(AB)} \epsilon^{R'S'} \right] \epsilon^{RS} \nonumber \\
{} + h^{(A |S}{}_{M'}{}^{R'|} h^{B)R M'S'} +h^{(A|R|} h^{B)S} \epsilon^{R'S'} 
+ h^{(A|N|} h^{B)}{}_{N}{}^{R'S'} \epsilon^{RS} \, ,
\end{align}
where the first line contains terms of degree zero and one in the tetrad perturbation, and the second line contains the quadratic terms. When contracted with the $\partial\omega$ part of the curvature, the 
first line gives rise to the kinetic term. The second term in the first line exhibits the combination $h^{AB}{}^{R'S'} - h^{(AB)} \epsilon^{R'S'}$ that motivates the introduction of the new field $H^{ABA'B'}$ in (\ref{H}).

\end{fmffile}

\end{document}